\documentclass[twocolumn,prstab,floatfix]{revtex4}
\usepackage{graphicx}
\usepackage{amsmath}
\usepackage{latexsym}
\usepackage{amsfonts}
\usepackage{amssymb}

\begin{document}

\title{Compensation of wake-field-driven energy spread in Energy Recovery Linacs}
\author{Georg H.~Hoffstaetter}
\author{Yang Hao Lau}
\affiliation{Cornell University, Ithaca, New York 14853}

\begin{abstract}

Energy Recovery Linacs provide high-energy beams, but decelerate those beams before dumping them, so that their energy is available for the acceleration of new particles. During this deceleration, any relative energy spread that is created at high energy is amplified by the ratio between high energy and dump energy. Therefore, Energy Recovery Linacs are sensitive to energy spread acquired at high energy, e.g. from wake fields. One can compensate the time-correlated energy spread due to wakes via energy-dependent time-of-flight terms in appropriate sections of an Energy Recovery Linac, and via high-frequency cavities. We show that nonlinear time-of-flight terms can only eliminate odd orders in the correlation between time and energy, if these terms are created by a beam transport within the linac that is common for accelerating and decelerating beams. If these two beams are separated, so that different beam transport sections can be used to produce time-of-flight terms suitable for each, also even-order terms in the energy spread can be eliminated. As an example, we investigate the potential of using this method for the Cornell x-ray Energy Recovery Linac. Via quadratic time-of-flight terms, the energy spread can be reduced by 66\%. Alternatively, since the energy spread from the dominantly resistive wake fields of the analysed accelerator is approximately harmonic in time, a high-frequency cavity could diminish the energy spread by 81\%. This approach would require bunch-lengthening and recompression in separate sections for accelerating and decelerating beams. Such sections have therefore been included in Cornell's x-ray Energy Recovery Linac design.

\end{abstract}

\maketitle

\section{Introduction}

Energy Recovery Linacs (ERLs) accelerate high-current particle beams to high energy in a linac. These are then used in x-ray \cite{Cornell1,Cornell2}, FEL \cite{Jlab1} or nuclear physics \cite{BNL1,BNL2} experiments. Subsequently, the beams are sent into the same linac at a decelerating phase to recover the particles' energy. This energy is then used to accelerate new bunches of particles \cite{Tigner65}. Only with such energy-recycling does it become feasible to accelerate high-current beams to high energies in a linac. Today's high-current, high-energy beams are produced by storage rings, which reuse energetic electrons for millions of turns. The beam emittance is then limited by the equilibrium emittance that is established during these turns. ERLs use each bunch of electrons only once, and therefore have the potential of producing significantly smaller emittances.

One of the problems ERLs face is the wake-field-driven energy spread that builds up during a pass through the ERL. The energy spread at high velocity is relatively small, but relevant because it limits the bandwidth of the x-ray radiation. More importantly, it is multiplied during deceleration by the ratio between high energy and dump energy, which is approximately 500 in the case of the Cornell ERL. Decelerated bunches must be decoupled from accelerated bunches, via magnetic fields, in a demerger region before the dump. Here, particles with too large energy are not bent sufficiently to reach the dump; similarly, particles with too little energy hit the beam pipe before the dump, or are even decelerated to zero energy before leaving the superconducting linac, and are then lost in the cryogenic environment. Therefore, the energy spread that bunches have at the end of their deceleration has to be limited.

\begin{figure}
\includegraphics[width=0.5\textwidth]{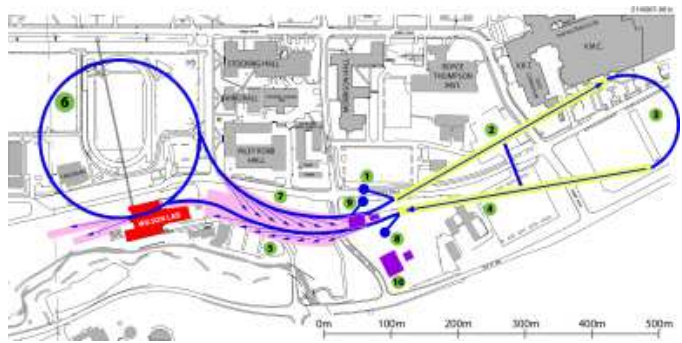}
\caption{Cornell ERL layout. 1) Injector, 2) linac~A, 3) turn-around loop, 4) linac~B, 5) south x-ray beamlines, 
6) CESR, 7) north x-ray beamlines, 8) first beam dump, 9) second beam dump and 10) distributed cryoplant.}
\label{fg:layout} 
\end{figure}

To illustrate the proposed techniques of wake-field compensation, we use the Cornell ERL, shown in Fig.~\ref{fg:layout}. Bunches enter the ERL via the injector at (1). They are accelerated first by linac~A at (2) and then by linac~B at (4) after traversing the turn-around loop connecting the linacs at (3). They then traverse the Cornell Electron Storage Ring (CESR) at (6), producing x-rays in beamline sections (5) and (7). Returning to linac~A, the bunches are decelerated, passed through the turn-around loop, and further decelerated by linac~B. Finally, the demerger leads decelerated bunches into the beam dump at (8).

\begin{figure}
\includegraphics[width=0.5\textwidth]{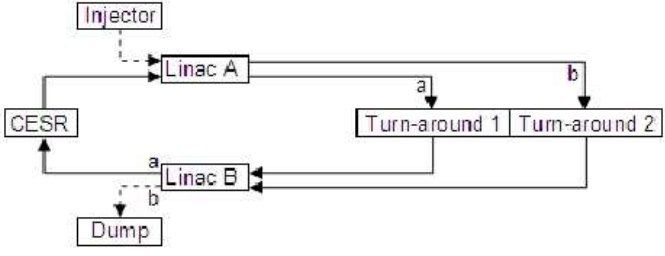}
\caption{Schematic of the Cornell ERL.}
\label{fg:schematic} 
\end{figure}

One can describe the propagation of a bunch's longitudinal phase-space distribution through the ERL via transformations for separate ERL components. In the Cornell ERL, shown schematically in Fig.~\ref{fg:schematic}, the bunch center is to be the design particle with $t_0=0$ and energy $E_0^c$. The bunch first traverses linac~A, which has frequency $\omega$, mapping the initial longitudinal phase-space position $\{t_0, E_0\}$ for each particle to

\begin{align}
\label{e:model0}
\{t_1, E_1\} = \{t_0, E_0 + \Delta E_{\rm{A}} \frac{\cos(\omega t_0 + \phi_{\rm{A}})}{\cos( \phi_{\rm{A}})}\}\ .
\end{align}

\noindent Here, the time coordinate after linac~A is chosen so that the bunch center arrives at $t_1=0$; similarly, we choose $t_i=0$ at the bunch center for each other section. $\Delta E_{\rm{A}}$ is the energy the linac adds to the bunch center. To limit the peak accelerating voltage $\frac{\Delta E_{\rm{A}}}{e \cos( \phi_{\rm{A}})}$, $\phi_{\rm{A}}$ will be kept below $15^o$. Next, the particles traverse the first turn-around loop with a time-of-flight mapping

\begin{align}
\label{e:model1}
\{t_2, E_2\} &= \{t_1 + T56_{\rm{TA1}} (E_1 - E_1^c) \\
&+ T566_{\rm{TA1}} (E_1 - E_1^c)^2, E_1\}\ , \nonumber
\end{align}

\noindent where the energy of the bunch center is $E_1^c = E_0^c + \Delta E_{\rm{A}}$. Subsequently, linac~B applies another accelerating field of frequency $\omega$ to the bunch, mapping $\{t_2, E_2\}$ to

\begin{align}
\label{e:model2}
\{t_3, E_3\} = \{t_2, E_2 + \Delta E_{\rm{B}} \frac{\cos(\omega t_2 + \phi_{\rm{B}})}{\cos(\phi_{\rm{B}})}\}\ .
\end{align}

\noindent Also $\phi_{\rm{B}}$ is limited to $15^o$. Next, we add half the accumulated effect of all wake fields to $E_3$:

\begin{align}
\label{e:model3}
\{t_4, E_4\} = \{t_3, E_3 + \frac{W(t_3)}{2}\}\ .
\end{align}

\noindent The particles then traverse CESR, with a time of flight mapping $\{t_4, E_4\}$ to

\begin{align}
\label{e:model4}
\{t_5, E_5\} &= \{t_4 + T56_{\rm{CE}} (E_4 - E_4^c) \\
&+ T566_{\rm{CE}} (E_4 - E_4^c)^2, E_4\}\ . \nonumber
\end{align}

\noindent Next, the second half of the accumulated effect of wake fields is added:

\begin{align}
\label{e:model5}
\{t_6, E_6\} = \{t_5, E_5 + \frac{W(t_5)}{2}\}\ .
\end{align}

\noindent Now, the bunch returns to linac~A, the second turn-around loop, and linac~B, which apply Eqs.~(\ref{e:model0}),~(\ref{e:model1})~and~(\ref{e:model2}) to yield

\begin{align}
\label{e:model6}
\{t_7, E_7\} = \{t_6, E_6 - \Delta E_{\rm{A}} \frac{\cos(\omega t_6 + \phi_{\rm{A}}')}{\cos(\phi_{\rm{A}})}\}\ ,
\end{align}

\noindent where $\phi_{\rm{A}}'=\phi_{\rm{A}} + \omega \delta t_{\rm{CE}}$,

\begin{align}
\label{e:model7}
\{t_8, E_8\} &= \{t_7 + T56_{\rm{TA2}} (E_7 - E_7^c) \\
&+ T566_{\rm{TA2}} (E_7 - E_7^c)^2, E_7\}\ , \nonumber
\end{align}

\begin{align}
\label{e:model8}
\{t_9, E_9\} = \{t_8, E_8 - \Delta E_{\rm{B}} \frac{\cos(\omega t_8 + \phi_{\rm{B}}')}{\cos(\phi_{\rm{B}})}\}\ ,
\end{align}

\noindent with $\phi_{\rm{B}}'=\phi_{\rm{B}}+\omega(\delta t_{\rm{CE}} + \delta t_{\rm{TA2}})$. In this state, the bunch leaves the ERL for the dump. While energy recovery of $\Delta E_{\rm{A}}$ in linac~A demands that the bunch center return to linac~A after an odd multiple of half the RF period, a possible deviation $\delta t_{\rm{CE}}$ is included above. Similarly, the decelerated beam might require a slightly different time than the accelerated beam to pass from linac~A to linac~B, which is described by $\delta t_{\rm{TA2}}$.

\section{Wake fields in the Cornell ERL}

\begin{figure}
\includegraphics[width=0.5\textwidth]{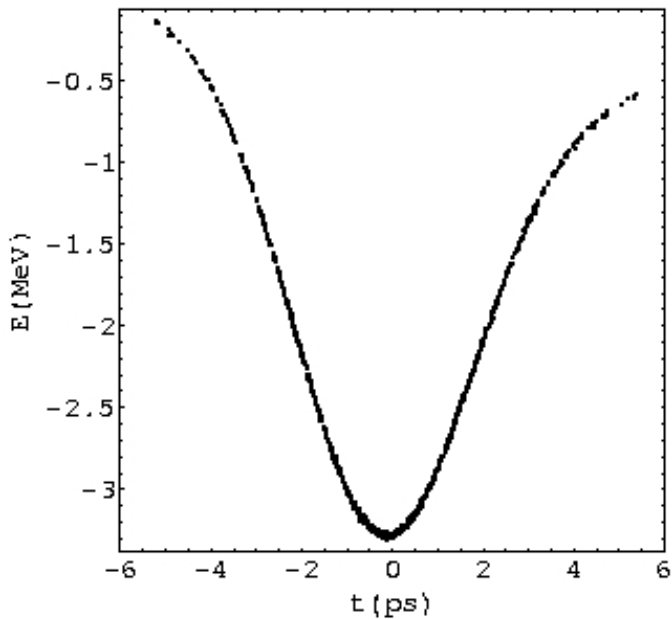}
\caption{Accumulated longitudinal wake potential of the Cornell ERL.}
\label{fg:wake} 
\end{figure}

\begin{figure}
\includegraphics[width=0.5\textwidth]{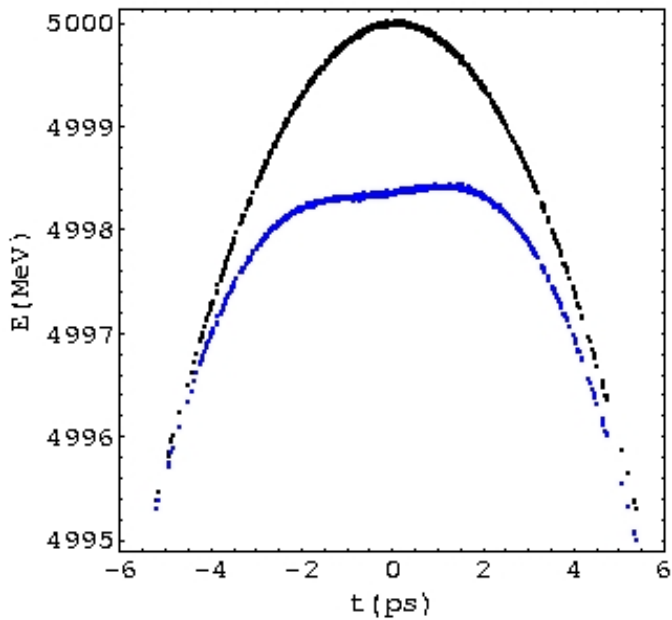}
\caption{Wake-induced bunch profiles at CESR. Black-top: Cosine-like correlated longitudinal phase space from accelerating on crest with a $\sigma_t=2$ps bunch length. Blue-bottom: Longitudinal profile after suffering half the Cornell ERL's wake field, $\frac{W(t)}{2}$.}
\label{fg:wake2} 
\end{figure}

Table~\ref{t:wake} lists the sources and magnitudes of the wake-induced energy spread for ERL components. The structure of $W(t)$ and its effect on the bunch at CESR for the Cornell ERL are shown in Figs.~\ref{fg:wake}~and~\ref{fg:wake2} respectively.

\begin{table}[htb!]
\begin{center}
\begin{tabular}{|c|c|c|}
\hline
Source  & Number & Max (kV/pC) \\
\hline
7 Cell RF Cavity	&	800	&	-11.32		\\
Higher Mode Load (78 mm)	&	400	&	-0.89		\\
Higher Mode Load (106 mm)	&	400	&	-0.50		\\
Expansion Joint	&	356	&	-0.74	\\
Beam Position Monitor (Button)	&	664	&	-0.35		\\
Beam Position Monitor (Stripline)	&	20	&	-0.01		\\
Flange Joint	&	356	&	-0.90		\\
Clearing Electrode	&	150	&	-0.18		\\
Gate Valve	&	68	&	-0.71		\\
1 m Resistive Wall (12.7 mm)	&	2500	&	-4.00		\\
1 m Roughness (12.7 mm)	&	2500	&	-14.00		\\
Undulator Taper (3 mm)	&	18	&	-0.61		\\
1 m Resistive Wall (3 mm)	&	144	&	-0.98		\\
1 m Roughness (3 mm)	&	144	&	-3.60		\\
\hline
\end{tabular}
\caption{Sources and magnitudes of wake-driven energy spread \cite{Billing}. Listed magnitudes are to be multiplied by the bunch charge of 77pC.}
\label{t:wake}
\end{center}
\end{table}

\section{Compensation methods for wake driven energy spread}

\subsection{Time of flight for wake correction}

\begin{figure}
\includegraphics[width=0.5\textwidth]{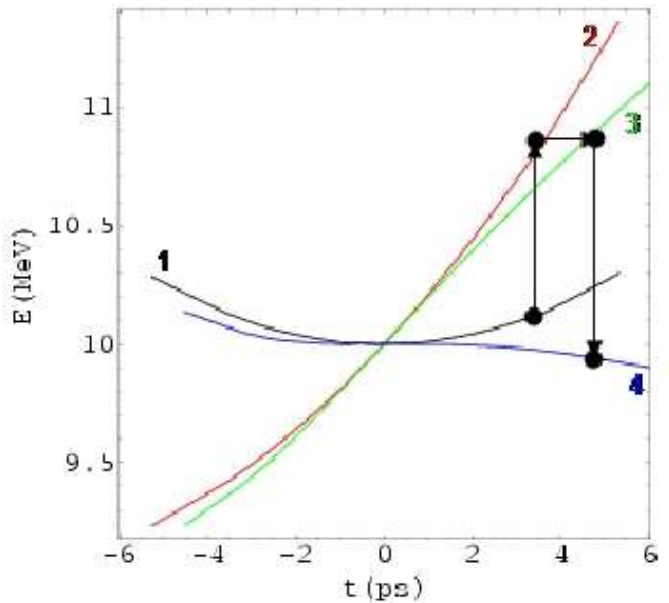}
\caption{Illustration of curvature change in $E(t)$ by off-crest acceleration and time-of-flight terms.}
\label{fg:visualization} 
\end{figure}

A bunch's correlated energy spread can be reduced by decreasing its slope and curvature in time-energy phase space. This can be done by accelerating the bunch off-crest in the linacs, choosing an energy-dependent time of flight, and decelerating the bunch off-crest. We sketch this procedure in Fig.~\ref{fg:visualization}. There, arrows show the phase-space motion of a particle. Curve 1 shows the initial bunch profile; curve 2 shows the bunch after acceleration by the linacs; curve 3 shows the bunch after application of time-of-flight terms; curve 4 shows the decelerated bunch, where the bunch's initial curvature, i.e. second-order correlation between time and energy, is eliminated. The arrows follow one particle of the bunch through this process.

\subsubsection{Linear time of flight for linear-wake correction}
\label{s:linear}

We first eliminate the average slope $\frac{\Delta E_9}{T} = \frac{E_9(\frac{T}{2})-E_9(-\frac{T}{2})}{T}$ of the phase-space bunch profile before the dump, where $T$ is chosen to be six times as large as $\sigma_t$, the rms temporal bunch length. Eliminating $\Delta E_9$ may not necessarily decrease the energy spread, but serves to symmetrize the wake for higher-order compensation. We minimize $|\Delta E_9|$ using time-of-flight terms. Additionally, $T56_{\rm{TA1}}$ can be used to minimize the energy spread $|\Delta E_4|$ in CESR. This minimization has to be done numerically, and we start with the $T56_{\rm{TA2}}$ that eliminates the phase-space slope $\frac{d E_9}{d t_9}$ at the bunch center before the dump.

We obtain $\frac{d E_9}{d t_9}$ from

\begin{align}
\label{e:slope}
\frac{d E_i}{d t_i}=\frac{\frac{d E_i}{d t_0}}{\frac{d t_i}{d t_0}}\ ,
\end{align}

\noindent with $\frac{d t_i}{d t_0}$ and $\frac{d E_i}{d t_0}$ from

\begin{align}
&\vec{z}_i \equiv
\left( \begin{array}{c}
 t_i \\
 E_i
 \end{array} \right)\ , \quad
\label{e:linTrans}
&\frac{d \vec{z}_i}{dt_0} = \underline{M}_{i0} \frac{d \vec{z}_0}{dt_0} = \underline{M}_{i0}
\left( \begin{array}{c}
 1 \\
 0
 \end{array} \right)\ ,
\end{align}

\noindent where we have chosen $\frac{dE_0}{dt_0}=0$ for phase-space distributions that enter the linac without linear time-energy correlation. Here,

\begin{align}
\underline{M}_{ij} = \prod_{k=i}^{j+1}{\underline{M}_{kk-1}}\ ; \quad \underline{M}_{kk-1} = \frac{\partial (t_{k}, E_{k})}{\partial(t_{k-1}, E_{k-1})}
\end{align}

\noindent describes the transfer matrices in longitudinal phase space;

\begin{align}
\underline{M}_{10}=
\left( \begin{array}{cc}
1 & 0\\
-\omega \Delta E_{\rm{A}}\tan(\phi_{\rm{A}}) & 1
 \end{array} \right)\ ,
\end{align}

\begin{align}
\underline{M}_{21}=
\left( \begin{array}{cc}
1 & T56_{\rm{TA1}} \\
0 & 1
 \end{array} \right)\ ,
\end{align}

\begin{align}
\underline{M}_{32}=
\left( \begin{array}{cc}
1 & 0 \\
-\omega \Delta E_{\rm{B}}\tan(\phi_{\rm{B}}) & 1
 \end{array} \right)\ ,
\end{align}

\begin{align}
\underline{M}_{43}=\underline{M}_{65}=
\left( \begin{array}{cc}
1 & 0 \\
\frac{W'(0)}{2} & 1
 \end{array} \right)\ ,
\end{align}

\begin{align}
\underline{M}_{54}=
\left( \begin{array}{cc}
1 & T56_{\rm{CE}} \\
0 & 1
 \end{array} \right)\ ,
\end{align}

\begin{align}
\underline{M}_{76}=
\left( \begin{array}{cc}
1 & 0 \\
\omega \Delta E_{\rm{A}}\frac{\sin(\phi_{\rm{A}}')}{\cos(\phi_{\rm{A}})} & 1
 \end{array} \right)\ ,
\end{align}

\begin{align}
\underline{M}_{87}=
\left( \begin{array}{cc}
1 & T56_{\rm{TA2}} \\
0 & 1
 \end{array} \right)\ ,
\end{align}

\begin{align}
\underline{M}_{98}=
\left( \begin{array}{cc}
1 & 0 \\
\omega \Delta E_{\rm{B}}\frac{\sin(\phi_{\rm{B}}')}{\cos(\phi_{\rm{B}})} & 1
 \end{array} \right)\ .
\end{align}

\subsubsection{Nonlinear time of flight for nonlinear-wake correction}

With the average phase-space slope of the bunch profile before the dump eliminated, the next step is to reduce the phase-space curvature $\frac{d^2E_9}{dt_9^2}$ at the bunch center before the dump. Unfortunately, minimizing the absolute curvature at the bunch center does not necessarily minimize the total energy spread. Therefore, we again use numerical minimization, which we start with analytically determined parameters that eliminate $\frac{d^2E_9}{dt_9^2}$ with $\underline{J} = 
\left( \begin{array}{cc}
0 & 1 \\
-1 & 0
 \end{array} \right)$,

\begin{align}
\label{e:curvature}
\frac{d^2E_i}{dt_i^2} &= \frac{\frac{d^2E_i}{dt_0^2}\frac{dt_i}{dt_0} - \frac{dE_i}{dt_0} \frac{d^2t_i}{dt_0^2}}{\frac{dt_i}{dt_0}^3} \\
&= \Big(\frac{dt_i}{dt_0}\Big)^{-3}\frac{d \vec{z}_i^T}{dt_0} \underline{J} \frac{d^2 \vec{z}_i}{dt_0^2}
= \Big(\frac{dt_i}{dt_0}\Big)^{-3}\frac{d \vec{z}_0^T}{dt_0} \underline{M}_{i0}^{T} \underline{J} \frac{d^2 \vec{z}_i}{dt_0^2}\ . \nonumber
\end{align}

\noindent The following iteration determines $\frac{d^2 \vec{z}_i}{dt_0^2}$;

\begin{align}
\label{e:iteration}
&\frac{d^2 \vec{z}_i}{dt_0^2} =  \vec{g}_i + \underline{M}_{ii-1} \frac{d^2 \vec{z}_{i-1}}{dt_0^2} ~\rm{with} \\
&\vec{g}_i = \frac{d \vec{z}_{i-1}^T}{dt_0} \vec{\underline{H}}_i \frac{d \vec{z}_{i-1}}{dt_0}\ , \\
 &\vec{\underline{H}}_i = \left( \begin{array}{c}
 \underline{H}[t_{i}(t_{i-1},E_{i-1})] \\
 \underline{H}[E_{i}(t_{i-1},E_{i-1})]
 \end{array} \right)\ ,
\end{align}

\noindent where $\underline{H}[f(x,y)]$ is the Hessian matrix of $f(x,y)$;

\begin{align}
\underline{H}[f(x,y)] =
\left( \begin{array}{cc}
\frac{\partial^2 f}{\partial x^2} & \frac{\partial^2 f}{\partial x \partial y} \\
\frac{\partial^2 f}{\partial y \partial x} & \frac{\partial^2 f}{\partial y^2}
\end{array} \right)\ ,
\end{align}

\noindent and the $\frac{d \vec{z}_{i}}{dt_0}$ are determined by Eq.~(\ref{e:linTrans}). Equation~(\ref{e:iteration}) leads to

\begin{align}
\label{e:curvatureForm0}
&\frac{d^2 \vec{z}_i}{dt_0^2} =  \sum_{k=1}^{i}{\underline{M}_{ik} \vec{g}_k} + \underline{M}_{i0} \left( \begin{array}{c}
0 \\
\frac{d^2 E_0}{dt_0^2}
\end{array} \right)\ ,
\end{align}

\noindent with $\frac{d^2 E_0}{dt_0^2}$ being the curvature in longitudinal phase space with which the bunches enter the linac. We here assume that the bunch enters the ERL without time-energy correlation; i.e. $\frac{dE_0}{dt_0} = \frac{d^2 E_0}{dt_0^2} = 0$. With Eq.~(\ref{e:curvature}) and using the symplecticity of transport matrices, i.e. $\underline{M}^T\underline{J}~\underline{M}=\underline{J}$, Eq.~(\ref{e:curvatureForm0}) leads to

\begin{align}
\label{e:curvatureForm}
\frac{d^2E_i}{dt_i^2} = \Big(\frac{dt_i}{dt_0}\Big)^{-3}(1,0)\sum_{k=1}^i{\underline{M}_{k0}^{T} \underline{J}\vec{g}_k}\ .
\end{align}

\noindent The Hessian matrices evaluated at the bunch center are zero except for

\begin{align}
H[E_1(t_0,E_0)]=
\left( \begin{array}{cc}
-\omega^2 \Delta E_{\rm{A}} & 0 \\
0 & 0
 \end{array} \right)\ ,
\end{align}

\begin{align}
H[t_2(t_1,E_1)]=
\left( \begin{array}{cc}
0 & 0 \\
0 & 2 T566_{\rm{TA1}}
 \end{array} \right)\ ,
\end{align}

\begin{align}
H[E_3(t_2,E_2)]=
\left( \begin{array}{cc}
-\omega^2 \Delta E_{\rm{B}} & 0 \\
0 & 0
 \end{array} \right)\ ,
\end{align}

\begin{align}
H[E_4(t_3,E_3)]=H[E_6(t_5,E_5)]=
\left( \begin{array}{cc}
\frac{W''(0)}{2} & 0 \\
0 & 0
 \end{array} \right)\ ,
\end{align}

\begin{align}
H[t_5(t_4,E_4)]=
\left( \begin{array}{cc}
0 & 0 \\
0 & 2 T566_{\rm{CE}}
 \end{array} \right)\ ,
\end{align}

\begin{align}
H[E_7(t_6,E_6)]=
\left( \begin{array}{cc}
\omega^2 \Delta E_{\rm{A}} \frac{\cos(\phi_{\rm{A}}')}{\cos(\phi_{\rm{A}})} & 0 \\
0 & 0
 \end{array} \right)\ ,
\end{align}

\begin{align}
H[t_8(t_7,E_7)]=
\left( \begin{array}{cc}
0 & 0 \\
0 & 2 T566_{\rm{TA2}}
 \end{array} \right)\ ,
\end{align}

\begin{align}
H[E_9(t_8,E_8)]=
\left( \begin{array}{cc}
\omega^2 \Delta E_{\rm{B}} \frac{\cos(\phi_{\rm{B}}')}{\cos(\phi_{\rm{B}})} & 0 \\
0 & 0
 \end{array} \right)\ .
\end{align}

\subsubsection{CESR time of flight}

As explained in Fig.~\ref{fg:visualization}, a first-order energy correlation has to be admitted in CESR to influence the curvature of the longitudinal distribution. Consequently, the bunch entering CESR has a large energy spread, which undesirably broadens the bunch's x-ray spectrum, rendering it unfeasible to use time-of-flight terms in CESR for wake compensation.

\subsubsection{Common turn-around loop for two ERL beams}

In this section, we choose  $\phi_{\rm{A}}' = \phi_{\rm{A}}$ and $\phi_{\rm{B}}' = \phi_{\rm{B}}$ for balanced acceleration and deceleration in each linac. Then, for the reference energy of the turn-around loop to be the same for both accelerating and decelerating beams, $E_1^c=E_7^c$, one has to choose $\Delta E_{\rm{A}} = \Delta E_{\rm{B}} + W(0)$, or for a total acceleration of $\Delta E = \Delta E_{\rm{A}} + \Delta E_{\rm{B}}$, $\Delta E_{\rm{A}} = \frac{\Delta E + W(0)}{2}$ and  $\Delta E_{\rm{B}} = \frac{\Delta E - W(0)}{2}$.

The slope, $\frac{dE_4}{dt_4}$, at the bunch center in CESR is to be eliminated to have small energy spread for x-ray experiments. Consequently, $\frac{dE_4}{dt_4}=0$ and

\begin{align}
\frac{d \vec{z}_7}{d t_0}
=\underline{M}_{74}
\left( \begin{array}{c}
\frac{dt_4}{dt_0} \\
0
 \end{array} \right)
=\frac{dt_4}{dt_0}
\left( \begin{array}{c}
1 \\
\omega \Delta E_{\rm{A}} \tan(\phi_{\rm{A}})+\frac{W'(0)}{2}
 \end{array} \right)\ ,
\end{align}

\noindent so Eq. (\ref{e:slope}) gives the slope at the bunch center in the second pass through the turn-around loop as $\frac{dE_7}{dt_7}=\omega \Delta E_{\rm{A}} \tan(\phi_{\rm{A}})+\frac{W'(0)}{2}$. Eq. (\ref{e:slope}) also gives the slope at the bunch center in the first pass through the turn-around loop, $\frac{dE_1}{dt_1}=-\omega \Delta E_{\rm{A}} \tan(\phi_{\rm{A}})$, and the beam therefore enters the turn-around loop with roughly opposite slopes on each pass; $\frac{dE_1}{dt_1} \simeq -\frac{dE_7}{dt_7}$.

\begin{figure}
\includegraphics[width=0.5\textwidth]{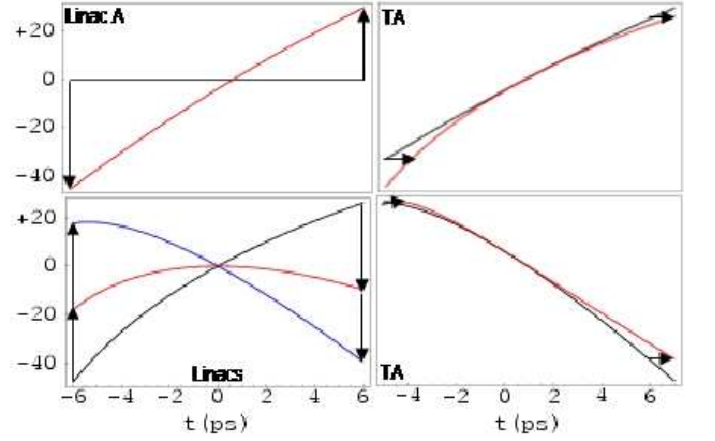}
\caption{Illustration of curvature cancellation in $E(t)$ by time-of-flight terms in a common turn-around loop for accelerating and decelerating beams. The ordinate specifies the energy relative to the bunch center in MeV.}
\label{fg:TA} 
\end{figure}

As a result, the net curvature change in the turnaround is zero, as explained in Fig.~\ref{fg:TA}. First, linac~A accelerates off-crest to produce a correlated energy spread here with positive slope. Second, the turn-around loop adds a second order time-of-flight shift, here to the right and thus effectively to the bottom of the bunch. In the second pass through the turn-around loop, the slope is negative, so that the second order time-of-flight terms shift effectively to the top of the bunch, compensating the curvature change in the first pass. The curvature from the first turn-around loop is counterbalanced by the curvature from the second turn-around loop. Evidently, this method cannot change even-order time-energy correlations. Only odd-order time-energy correlations at the dump can be eliminated when $\frac{dE_4}{dt_4}$ is to be eliminated.

\bigskip

\subsubsection{Separate turn-around loops for each ERL beam.}

In the previous two sections, we have seen how longitudinal phase-space curvature reduction can diminish the bunch energy spread, but can be feasibly carried out in neither CESR nor a single turn-around loop. However, passing the bunch through different turn-around loops before and after CESR enables such curvature reduction.

In this section, we choose  $\phi_{\rm{A}}' = \phi_{\rm{A}}$ and $\phi_{\rm{B}}' = \phi_{\rm{B}}$ for balanced acceleration and deceleration in each linac. The variable $T56_{\rm{TA1}}$ is chosen to eliminate the slope, $\frac{d E_4}{d t_4}$ from Eq.~(\ref{e:slope}), at the bunch center in CESR;

\begin{align}
\label{e:T56TA1}
T56_{\rm{TA1}} &= [\omega \Delta E_{\rm{A}} \tan(\phi_{\rm{A}})]^{-1}\\
&- [\frac{W'(0)}{2} - \omega \Delta E_{\rm{B}} \tan(\phi_{\rm{B}})]^{-1}\ ,\nonumber
\end{align}

\noindent and $T56_{\rm{TA2}}$ is chosen to eliminate $\frac{d E_9}{d t_9}$ at the dump;

\begin{align}
\label{e:T56TA2}
T56_{\rm{TA2}} &= -[\omega \Delta E_{\rm{B}} \tan(\phi_{\rm{B}})]^{-1}\\
&- [\frac{W'(0)}{2} + \omega \Delta E_{\rm{A}} \tan(\phi_{\rm{A}})]^{-1}\ .\nonumber
\end{align}

\noindent Similarly, $T566_{\rm{TA1}}$ will be used to eliminate the curvature, $\frac{d^2 E_4}{d t_4^2}$ from Eq.~(\ref{e:curvatureForm}), at the bunch center in CESR, and $T566_{\rm{TA2}}$ to eliminate that at the dump, leading to

\begin{align}
\label{e:T566TA1}
T566_{\rm{TA1}} &= \frac{2W''(0) - 4 \omega^2 \Delta E_{\rm{B}}}{[2\omega \Delta E_{\rm{B}} \tan(\phi_{\rm{B}})-W'(0)]^3}\\
 &+ \frac{\Delta E_{\rm{A}}}{2\omega[\Delta E_{\rm{A}} \tan(\phi_{\rm{A}})]^3}\ ,\nonumber
\end{align}

\begin{align}
\label{e:T566TA2}
T566_{\rm{TA2}} &= \frac{2 W''(0) + 4 \omega^2 \Delta E_{\rm{A}}}{[2\omega \Delta E_{\rm{A}} \tan(\phi_{\rm{A}})+W'(0)]^3}\\
 &- \frac{\Delta E_{\rm{B}}}{2\omega[\Delta E_{\rm{B}} \tan(\phi_{\rm{B}})]^3}\ .\nonumber
\end{align}

\noindent This choice compensates the curvature of the wake, as well as the curvature of the RF cosine wave. For the Cornell ERL, $W'(0) \simeq 0$, simplifying Eqs.~(\ref{e:T566TA1})~and~(\ref{e:T566TA2}). The part in Eqs.~(\ref{e:T566TA1})~and~(\ref{e:T566TA2}) that remains when setting $W'(0)=W''(0)=0$ corrects the curvature of the acceleration and deceleration cosine function \cite{Bazarov}; the rest is responsible for correcting the curvature of the wake fields. The phases $\phi_{\rm{A}}$ and $\phi_{\rm{B}}$ can be chosen independently. Subsequently we choose $\phi_{\rm{B}}$ so that $T56_{\rm{TA1}}=T56_{\rm{TA2}}=0$ when no wake field is present; i.e.
$\Delta E_{\rm{A}} \tan(\phi_{\rm{A}})=\Delta E_{\rm{B}} \tan(\phi_{\rm{B}})$.

\begin{figure}
\includegraphics[width=0.5\textwidth]{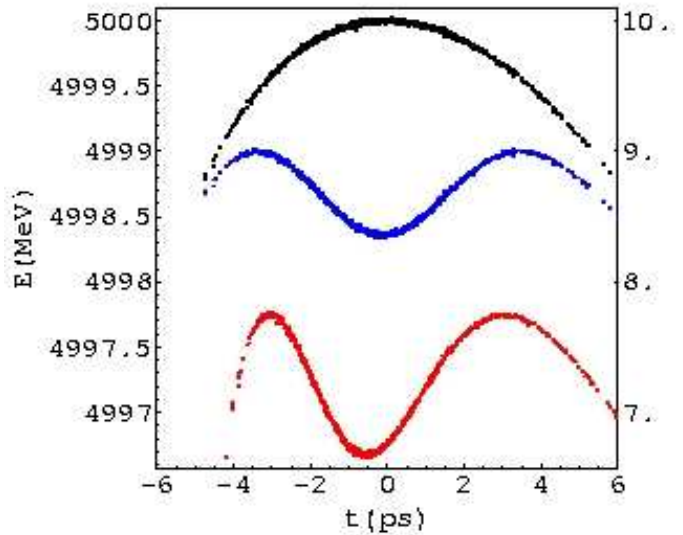}
\caption{Bunch profiles after nonlinear-wake correction with separate turn-around loops for each ERL beam. Left ordinate gives energy at CESR while right ordinate gives energy at dump. Black-top: Cosine-like correlated longitudinal phase space from accelerating on crest with a $\sigma_t=2$ps bunch length. Blue-middle: Longitudinal profile after suffering half the Cornell ERL's wake field. Red-bottom: Longitudinal profile at dump with residual energy spread due to higher-order correlations.}
\label{fg:2_TA} 
\end{figure}

We then numerically find the combination of $T56_{\rm{TA1}}$ and $T566_{\rm{TA1}}$ that minimizes the maximum energy difference between bunch particles at CESR, starting with Eqs.~(\ref{e:T56TA1})~and~(\ref{e:T566TA1}). Subsequently, we numerically find the combination of $T56_{\rm{TA2}}$ and $T566_{\rm{TA2}}$ that minimizes the maximum energy difference between bunch particles at the dump, starting with Eqs.~(\ref{e:T56TA2})~and~(\ref{e:T566TA2}). As an example we use $\Delta E_{\rm{A}}=\Delta E_{\rm{B}}=2495$MeV and $\phi_{\rm{A}}=-\phi_{\rm{B}}=15^o$. Figure~\ref{fg:2_TA} shows the best result, which has been obtained for $T56_{\rm{TA1}}=4.4\times10^{-5}$psMeV$^{-1}$, $T566_{\rm{TA1}}=7.7\times10^{-4}$psMeV$^{-2}$, $T56_{\rm{TA2}}=-5.6\times10^{-3}$psMeV$^{-1}$, $T566_{\rm{TA2}}=1.5\times10^{-3}$psMeV$^{-2}$ and $T56_{\rm{CE}}=T566_{\rm{CE}}=0$, with the maximum energy difference between bunch particles decreasing by $66\%$, from 3.2 to 1.1Mev. The residual energy spread is due to higher-order terms in $W(t)$. Hence, we acquire a large decrease in energy spread, without having had to send the bunch into CESR with an undesirable energy spread.

\subsection{Harmonic-wake correction}
\subsubsection{High-frequency cavity}
\label{s:harmonic}

Instead of attempting to minimize phase-space curvature at the bunch center before the dump by nonlinear time-of-flight terms, one could add energy to the bunch to flatten its longitudinal phase-space distribution. We simulate this by inserting a cavity with period $T$ approximately six times as large as $\sigma_t$, the rms temporal bunch length, immediately after CESR. $T$ is chosen to ensure the cavity is a multiple of the linac frequency, so subsequent bunches have the same phase at the cavity. The cavity maps the phase-space position after CESR, $\{t_6, E_6\}$ from Eq.~(\ref{e:model5}), to

\begin{align}
\label{e:model5'}
\{t_6', E_6'\} = \{t_6, E_6 + \Delta E_1 \cos(\frac{2 \pi [t_6-t_1^*]}{T})\}\ ,
\end{align}

\noindent so $t_6$ and $E_6$ in Eq.~(\ref{e:model6}) become $t_6'$ and $E_6'$ respectively.

In Eq.~(\ref{e:model5'}), $t_1^*$ and $\Delta E_1$ are determined from Fourier analysing $W(t)$:

\begin{align}
\label{e:Fourier}
&W(t) = \sum_{n=0}^{\infty} \Delta E_n \cos ( \frac{2 n \pi}{T} [t_5-t_n^*] )\ .
\end{align}

\noindent We again choose $\phi_{\rm{A}}' = \phi_{\rm{A}}$ and $\phi_{\rm{B}}' = \phi_{\rm{B}}$ for balanced acceleration and deceleration in each linac, and we choose $\phi_{\rm{B}}$ to satisfy Eq. (\ref{e:T56TA1}), which eliminates the slope at the bunch center in CESR.

\begin{figure}
\includegraphics[width=0.5\textwidth]{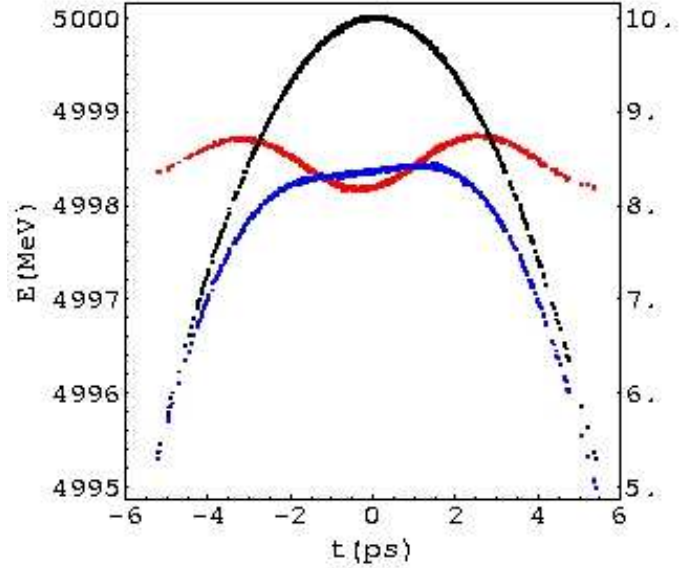}
\caption{Bunch profiles after harmonic-wake correction with high-frequency cavity. Left ordinate gives energy at CESR while right ordinate gives energy at dump. Black-top: Cosine-like correlated longitudinal phase space from accelerating on crest with a $\sigma_t=2$ps bunch length. Blue-bottom: Longitudinal profile after suffering half the Cornell ERL's wake field. Red-middle: Longitudinal profile at dump with residual energy spread due to higher harmonics.}
\label{fg:cavity} 
\end{figure}

Using a single turn-around loop, we numerically find the combination of $T56_{\rm{TA}}\equiv T56_{\rm{TA1}}\equiv T56_{\rm{TA2}}$, $t_1^*$ and $\Delta E_1$ that minimizes the maximum energy difference between bunch particles at the dump, starting with Eq.~(\ref{e:Fourier}). As an example we use $\Delta E_{\rm{A}}\simeq\Delta E_{\rm{B}}\simeq2495$MeV and $\phi_{\rm{A}}=10^o$, which is non-zero because there has to be a slope in longitudinal phase space to eliminate the first-order correlation via $T56_{\rm{TA}}$. Figure~{\ref{fg:cavity}} shows the best result, which has been obtained for $t_1^*=-9.0\times10^{-2}$ps, $\Delta E_1=1.5$MeV, $T56_{\rm{TA}}=-1.3\times10^{-3}$psMeV$^{-1}$, $\phi_{\rm{B}}=-9.9^o$, and $T566_{\rm{TA1}}=T566_{\rm{TA2}}=T56_{\rm{CE}}=T566_{\rm{CE}}=0$, with the maximum energy difference between bunch particles decreasing by $81\%$, from 3.2 to 0.60MeV. The residual energy spread is due to higher harmonics in $W(t)$. The decrease in energy spread is larger than that obtained using nonlinear-wake correction, indicating the effectiveness of harmonic correction. 

Although the decrease in energy spread is substantial, the required cavity frequency $\frac{2 \pi}{T} \simeq 80$GHz is too high to be economically feasible. The energy that has to be added to the bunch center is approximately 2MV, for a beam current of 0.1A. While power sources of up to 40GHz could be made available, a power of 200kW seems unfeasible. An X-band power sources at approximately 11.7GHz might be feasible, but the bunch length would have to be increased to do harmonic-wake correction with this lower frequency.

\subsubsection{Bunch expansion and recompression}
\paragraph{Common turn-around loop for two ERL beams.}

Reducing the frequency to 11.7GHz requires expanding the $6\sigma_t$ of the bunch by 7 from 12ps to 84ps. Because a longer bunch acquires more energy spread in the cosine-like RF field, we should recompress the bunch immediately after the 11.7GHz cavity. 

Resizing the bunch in CESR requires energy spread in CESR, and is therefore unfeasible. Using the turn-around loop allows for the compensation of energy spread due to $\phi_{\rm{A}}$ via a suitable $\phi_{\rm{B}}$. Here we simulate a cavity in the center of a single turn-around loop, with $T56_{\rm{TA}}$ describing the first half of the turn-around loop, to map the phase-space positions, $\{t_2,E_2\}$ and $\{t_8,E_8\}$ from Eqs.~(\ref{e:model1})~and~(\ref{e:model7}), to

\begin{align}
\label{e:model2'}
\{t_2', E_2'\} = \{t_2, E_2 - \Delta E_1 \cos(\frac{2 \pi [t_2-t_1^*]}{T})\}\ ,\\
\label{e:model8'}
\{t_8', E_8'\} = \{t_8, E_8 + \Delta E_1 \cos(\frac{2 \pi [t_8-t_1^*]}{T})\}\ ,
\end{align}

\noindent where $t_1^*$ and $\Delta E_1$ are determined from Eq. (\ref{e:Fourier}), with $W(t)$ replaced by the difference between $E_7$ and $E_7^{(0)}$, which is $E_7$ obtained from setting $W(t)=0$. Here, $T$ is approximately six times the new rms temporal bunch length. $T$ is chosen to ensure the cavity frequency is an odd multiple of the linac frequency, so the bunch has the opposite phase on both passes through the cavity. The second half of the turn-around loop maps $\{t_2', E_2'\}$ and $\{t_8', E_8'\}$ to

\begin{align}
\label{e:model2''}
\{t_2'', E_2''\} &= \{t_2' + T56'_{\rm{TA}} (E'_2 - E'^{c}_2) \\
&+ T566'_{\rm{TA}} (E_2' - E'^{c}_2)^{2}, E_2'\}\ , \nonumber \\
\label{e:model8''}
\{t_8'', E_8''\} &= \{t_8' + T56'_{\rm{TA}} (E_8' - E'^{c}_8) \\
&+ T566'_{\rm{TA}} (E_8' - E'^{c}_8)^{2}, E_8'\}\ , \nonumber
\end{align}

\noindent where we choose $T56'_{\rm{TA}} = -T56_{\rm{TA}}$ to correct the bunch length change from the first half of the turn-around loop. The numerical minimization will lead to a deviation of these two time-of-flight terms in order to find the smallest energy spread. For simplicity, we assume $T566'_{\rm{TA}} = -T566_{\rm{TA}}=0$, and do not use time-of-flight terms in CESR. Eqs.~(\ref{e:model2''})~and~(\ref{e:model8''}) imply that $t_i$ and $E_i$ in Eqs.~(\ref{e:model2})~and~(\ref{e:model8}) become $t_i''$ and $E_i''$ respectively, for $i=2, 8$.

We again choose $\phi_{\rm{A}}' = \phi_{\rm{A}}$ and $\phi_{\rm{B}}' = \phi_{\rm{B}}$ for balanced acceleration and deceleration in each linac, and $\phi_{\rm{B}}$ to eliminate the slope at the bunch center in CESR.

To determine the $T56_{\rm{TA}}$ suitable for this method, we obtain the expansion factor $\frac{dt_8}{dt_7} = \frac{\frac{dt_8}{dt_0}}{\frac{dt_7}{dt_0}}$, using $\frac{dt_8}{dt_0}$ and $\frac{dt_7}{dt_0}$ from Eq.~(\ref{e:linTrans}). Then, at the bunch center,

\begin{align}
\frac{dt_8}{dt_7} = 1 - T56_{\rm{TA}} \omega \Delta E_{\rm{B}}\tan(\phi_{\rm{B}})\ ,
\end{align}

\noindent where $W(t)$ and $\Delta E_1$ have been set to zero. To expand the decelerating beam by a factor of 7, we require $T56_{\rm{TA}} = -\frac{6}{\omega \Delta E_{\rm{B}}\tan(\phi_{\rm{B}})}$.

During the first pass through the turn-around loop, we obtain

\begin{align}
\frac{dt_2}{dt_1} = 1 - T56_{\rm{TA}} \omega \Delta E_{\rm{A}}\tan(\phi_{\rm{A}}) = -5\ ,
\end{align}

\noindent with $\Delta E_{\rm{A}}\tan(\phi_{\rm{A}}) = -\Delta E_{\rm{B}}\tan(\phi_{\rm{B}})$. The bunch is therefore expanded in the high-frequency cavity during both passes through the turn-around loop. 

\begin{figure}
\includegraphics[width=0.5\textwidth]{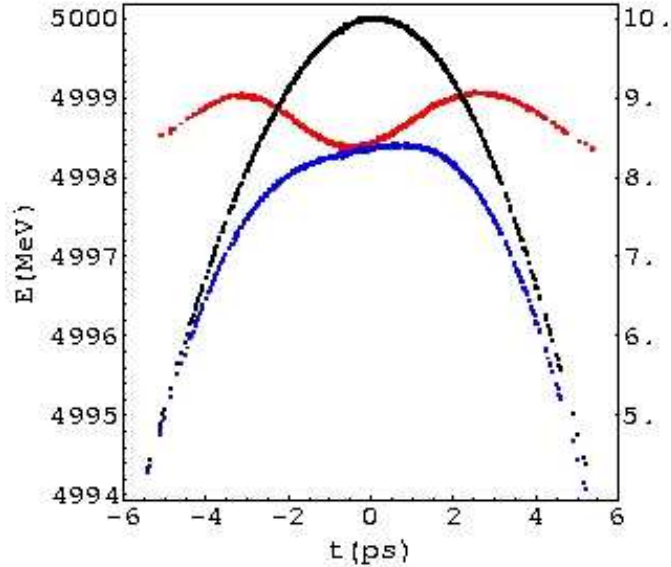}
\caption{Harmonic-wake correction with bunch expansion and recompression, in a single turn-around loop. Left ordinate gives energy at CESR while right ordinate gives energy at dump. Black-top: Cosine-like correlated longitudinal phase space from accelerating on crest with a $\sigma_t=2$ps bunch length. Blue-bottom: Longitudinal profile after suffering half the Cornell ERL's wake field. Red-middle: Longitudinal profile at dump.}
\label{fg:expansion} 
\end{figure}

We then numerically find the combination of $T56_{\rm{TA}'}$, $t_1^*$ and $\Delta E_1$ that minimizes the maximum energy difference between bunch particles at the dump, starting with Eq.~(\ref{e:Fourier}). As an example we use $\Delta E_{\rm{A}}\simeq\Delta E_{\rm{B}}\simeq2495$MeV, $\phi_{\rm{A}}=15^o$. Figure~{\ref{fg:expansion}} shows the best result, which was obtained for $t_1^*=5.9\times10^{-2}$ps, $\Delta E_1=0.14$MeV, $T56_{\rm{TA}'}=-1.1$psMeV$^{-1}$ and $\phi_{\rm{B}}=-15^o$, with the maximum energy difference between bunch particles decreasing by $78\%$, from 3.2 to 0.72MeV.

This result is decent, but the method is unfeasible because the bunch is over-compressed during the first pass, and therefore has to pass a point where it is fully compressed to very short length. At this point, Coherent Synchotron Radiation (CSR) damage to the beam can be very severe.

\bigskip

\paragraph{Separate turn-around loops for each ERL beam.}

\begin{figure}
\includegraphics[width=0.5\textwidth]{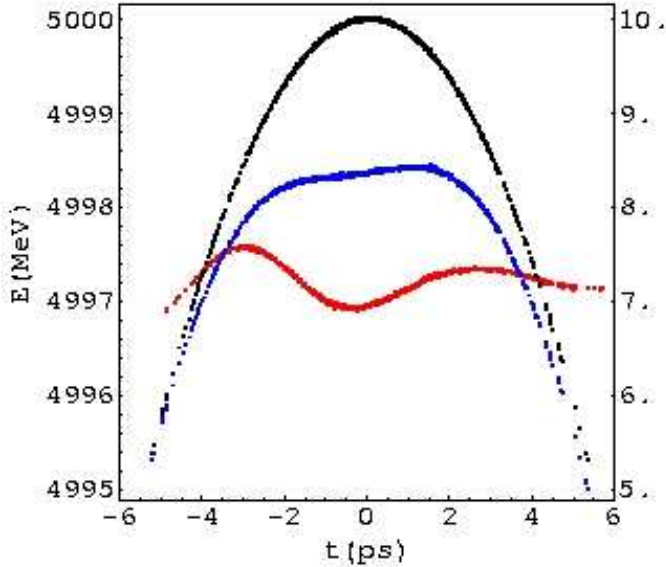}
\caption{Harmonic-wake correction with bunch expansion and recompression, in separate turn-around loops for each ERL beam. Left ordinate gives energy at CESR while right ordinate gives energy at dump. Black-top: Cosine-like correlated longitudinal phase space from accelerating on crest with a $\sigma_t=2$ps bunch length. Blue-middle: Longitudinal profile after suffering half the Cornell ERL's wake field. Red-bottom: Longitudinal profile at dump. Bunch profiles are similar to those in Fig.~\ref{fg:cavity} for harmonic-wake correction without bunch expansion and recompression.}
\label{fg:expansion_2TA} 
\end{figure}

The problem of over-compression and therefore very short bunches with strong CSR effect can be overcome by passing the bunches through different turn-around loops before and after CESR, such that only one of the turn-around loops carries out harmonic-wake correction. The harmonic correction is performed in the turn-around loop carrying the bunch after CESR, while the turn-around loop carrying the bunch before CESR has $T56_{\rm{TA1}}$ chosen according to Eq.~(\ref{e:T56TA1}) to eliminate the slope at the bunch center in CESR. The same expansion factor, $\frac{dt_8}{dt_7}=7$, and parameter choices, with the exceptions of $t_1^*=-0.61$ps, $\Delta E_1=0.21$MeV, allow us to decrease the maximum energy difference between bunch particles by $78\%$, from 3.2 to 0.69Mev. The result is shown in Fig.~{\ref{fg:expansion_2TA}}.

\section{Conclusion}

We have investigated the potential of using time-of-flight terms in ERL loops to reduce wake-driven correlated energy spread. As an example, we have used the Cornell ERL. One could use the time of flight in the ERL return pass that contains the x-ray sources, but this method requires undesirable correlated energy spread in this pass. We found that time-of-flight terms in the linac that are common to the accelerating and the decelerating beams cannot compensate even orders in the time-energy correlation. However, when the accelerating and the decelerating beams are separated in the linac, time of flight for the two passes through the linac can be used to reduce the energy spread in the return loop, as well as in the beam dump. The correlated energy spread is reduced by 66\% in our example.

Furthermore, we have looked into the potential of using high-frequency cavities to reduce the energy spread. We find that placing an 80Ghz cavity after CESR could eliminate the first harmonic of the energy spread, and decreases the energy variation by 81\% in our example. Unfortunately, a cavity with such a high frequency is unfeasible, necessitating bunch-lengthening to permit a lower cavity frequency. The method we have analysed involves expanding the bunch in the first half of the turn-around loop, correcting the first harmonic of the time-expanded energy spread in a cavity, then recompressing the bunch in the second half of the turn-around loop. Separate turn-around loops would be needed to avoid strong CSR forces during overcompression, yielding a 78\% decrease in energy variation in our example.

Since the wake-induced energy spread is better described by a sinusoid than a parabola, harmonic correction is somewhat more effective than nonlinear time-of-flight terms at minimizing the energy spread at the dump. But bunch compression and decompression, as well as a section with high frequency structures, are hard to devise. In both cases, two turn-around loops are required, and have thus been included in the Cornell ERL design.

\begin{figure}
\includegraphics[width=0.5\textwidth]{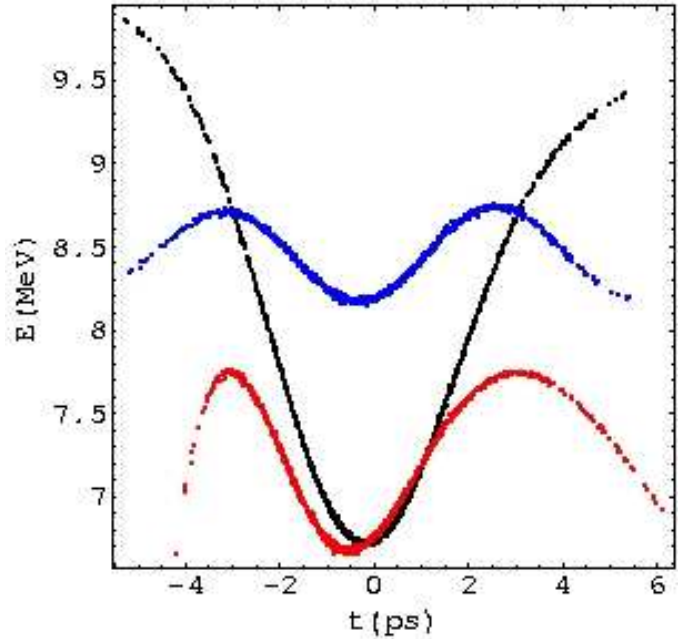}
\caption{Results. Black-top: Longitudinal profile at dump without wake-correction. Blue-middle: Dump profile with harmonic wake-correction. Red-middle: Dump profile with nonlinear time-of-flight wake-correction. Harmonic wake-correction reduces energy spread more but is less feasible than nonlinear wake-correction.}
\label{fg:results} 
\end{figure}

\subsection*{Acknowledgments}
We thankfully acknowledge useful discussions with Mike Billing, who has provided the wake potential for the Cornell ERL. This work has been supported by NSF cooperative agreement PHY-0202078.

\end{document}